\documentclass[12pt]{article}

\usepackage{epsf,array}

\evensidemargin\oddsidemargin
\def\ord#1{{\mathcal O}\l(#1\r)}

\def\btorho{\bar B^0\to\rho^+ l^- \bar\nu_l}
\def\btopi{\bar B^0\to\pi^+ l^- \bar\nu_l}
\def\btokstargamma{\bar B\to K^* \gamma}
\def\vub{|V_{ub}|}
\def\vts{|V_{ts}|}
\def\qsqmax{q^2_{\mathrm{max}}}

\def\dof{\mathrm{dof}}
\def\gev{\,\mathrm{Ge\kern-0.1em V}}
\def\mev{\,\mathrm{Me\kern-0.1em V}}
\newdimen\unit
\def\point#1 #2 #3{\vbox to0pt{\kern-#2\unit
  \hbox{\kern#1\unit$#3$}\vss}
 \nointerlineskip}

\def\be{\begin{equation}}
\def\ee{\end{equation}}
\def\bea{\begin{eqnarray}}
\def\eea{\end{eqnarray}}

\def\la{\langle}
\def\ra{\rangle}
\def\tr{\mathop{\mathrm{Tr}}\nolimits}
\makeatletter
\def\slash#1{{\mathpalette\c@ncel{#1}}} 
\makeatother
\def\l{\left}
\def\r{\right}

\def\eqs#1#2{Eqs.~(\ref{#1},\ref{#2})}
\def\eq#1{Eq.~(\ref{#1})}
\def\re#1{Ref.~\cite{#1}}

\def\sec#1{Section~\ref{#1}}
\def\fig#1{Fig.~\ref{#1}}
\def\figs#1#2{Figs.~\ref{#1} and~\ref{#2}}
\def\tab#1{Tab.~\ref{#1}}

\makeatletter
\newsavebox{\capbox}
\def\@makecaption#1#2{%
  \vspace{10pt}\sbox{\capbox}{\small\textbf{#1:} #2}%
  \hbox to\linewidth{\hfill
  \ifdim\wd\capbox > 0.9\linewidth%
    \parbox{0.9\linewidth}{\small\textbf{#1:} #2}%
   \else%
    \small\textbf{#1:} #2%
   \fi%
  \hfill}}
\makeatother

\renewcommand{\arraystretch}{1.2}

\newcommand{\err}[2]{%
{{\renewcommand{\arraystretch}{0.4}
\mathop{\raisebox{0.1\height}{\scriptsize
$\begin{array}{@{}c@{}}+\\-\end{array}$}}%
\raisebox{0.1\height}{\scriptsize
$\begin{array}{@{}r@{}}#1\\#2\end{array}$}}}%
}

\begin{document}

\begin{titlepage}
  
\renewcommand{\thefootnote}{\fnsymbol{footnote}}
  
\begin{flushright}
CPT--97/P.3505\\
SHEP--97/13\\
UG--DFM--4/97\\
hep-lat/9708008
\end{flushright}  
\bigskip

\begin{center}
\huge\bfseries
Lattice-Constrained Parametrizations of Form Factors for
Semileptonic and Rare Radiative $B$ Decays
\end{center}

\bigskip\bigskip

\begin{center}
{\bfseries UKQCD Collaboration}\\[1em]
{\bfseries Luigi Del Debbio$^1$, Jonathan M Flynn$^2$, Laurent
Lellouch$^1$ and Juan Nieves$^3$}\\[2em]
$^1$Centre de Physique Th\'eorique\footnote{
Unit\'e Propre de Recherche 7061}, CNRS Luminy, Case 907
F-13288 Marseille Cedex 9, France\\
$^2$Department of Physics and Astronomy, University of Southampton,
Highfield, Southampton SO17 1BJ, UK\\
$^3$Departamento de Fisica Moderna, Avenida Fuentenueva, 18071 Granada,
Spain
\end{center}

\vfill

\begin{center}  
{\bf Abstract}
\end{center}

We describe all the form factors for semileptonic and radiative $B$
decays to the same light vector meson with just two parameters and the
two form factors for semileptonic $B$ decays to a light pseudoscalar
meson with a further two or three parameters. This provides simple
parametrizations of the form factors for the physical $\btorho$,
$\btopi$ and $\btokstargamma$ decays. The parametrizations are
consistent with heavy quark symmetry, kinematic constraints and
lattice results, which we use to determine the parameters. In
addition, we test versions of the parametrizations consistent (or not)
with light-cone sum rule scaling relations at $q^2=0$.

\vfill



\setcounter{footnote}{0}
\renewcommand{\thefootnote}{\arabic{footnote}}
  
\end{titlepage}
  
\section{Motivation}

The aim of this note is to obtain a simple yet phenomenologically
useful description of the form factors for semileptonic and rare
radiative heavy-to-light meson decays for all $q^2$, the squared
four-momentum transfer to the leptons or photon. Lattice calculations
provide values for the form factors over a limited region at high
$q^2$. We supplement these data with further constraints and model
input to obtain our parametrization.

The semileptonic decays have been measured by
CLEO~\cite{cleo:semilept96}. Combining those results with our
parametrization leads to values for $\vub$. The decay $\btokstargamma$
depends on $\vts$ when described in the Standard Model, but since it
first occurs at one loop it is an excellent place to search for new
physics~\cite{mad1,mad2,hewett:topten,htt:ipnp}: in either case it is
important to know the relevant form factors.

Kinematic constraints relate some of the form factors at $q^2=0$, and
light cone sum rule (LCSR) calculations provide further information at
low $q^2$, particularly on the heavy meson mass dependence of the form
factors as that mass becomes large. Specifically, LCSR predict that
all the form factors scale like $m_B^{-3/2}$ in the $m_B \to\infty$
limit. At the other end of the range, near $\qsqmax$, heavy quark
symmetry (HQS) provides, at lowest order, extra relations between the
form factors, but (unlike the case of heavy-to-heavy transitions) does
not determine the overall normalisation. Extra assumptions are needed
to cover the full $q^2$ range.  For the decays with a light final
state vector meson, we apply a simple model due to
Stech~\cite{stech1}, which uses just two free parameters. For a light
final state pseudoscalar we use fits with two or three
parameters~\cite{lpl:btopi-bounds,ukqcd:hlff}. The normalisation in
each case is determined using lattice data from the UKQCD
collaboration~\cite{ukqcd:hlff,ukqcd:btorho,ukqcd:a0a3}.

\section{Description of Model}

In this section, we transcribe the assumptions made by Stech
in~\cite{stech1} into the language of HQS. For $q^2$ close to
$\qsqmax$, the standard leading order HQS analysis~\cite{iw:hqet,gmm}
for a $B$ meson decaying semileptonically or radiatively into a light
pseudoscalar ($\pi$) or vector ($\mathcal{V}$) meson allows the
corresponding matrix elements to be written as:
\be
\la \pi|\bar q\gamma^\mu b|\bar B\ra = \sqrt{m_\pi m_B} \tr\l\{
\l(\theta_1^\pi - \frac{\slash{p}_\pi}{m_\pi}\theta_2^\pi\r)\gamma^5\gamma^\mu
\frac{1+\slash{v}}{2}\gamma^5\r\}
\label{pseudo}
\ee
and
\bea
\la \mathcal{V}(\eta)|
\bar q\Gamma b|\bar B\ra = &-i \sqrt{m_\mathcal{V} m_B} \tr\l\{
\l[\l(\theta_1^\mathcal{V} + \frac{\slash{p}_\mathcal{V}}{m_\mathcal{V}}
\theta_2^\mathcal{V}\r)\slash{\eta}^*
 \r.\r.\nonumber\\
& + \l.\l.\l(\theta_3^\mathcal{V} + 
\frac{\slash{p}_\mathcal{V}}{m_\mathcal{V}}\theta_4^\mathcal{V}\r)v
\cdot\eta^*\r]\Gamma
\frac{1+\slash{v}}{2}\gamma^5\r\}
\ ,
\label{vector}
\eea
where $v$ is the four-velocity of the $B$ meson, $q$ is the light
active quark ($u$ or $s$), and $\Gamma$ denotes either
$\gamma^\mu(1-\gamma^5)$ or $q_\nu\sigma^{\mu\nu}(1+\gamma^5)/2$, with
$q_\nu = (p_B-p_{\pi,\mathcal{V}})_\nu$. The
$\theta_i^{\pi,\mathcal{V}}$ are functions of the invariant $v\cdot
p_{\pi,\mathcal{V}}$ and are independent of heavy-quark mass and spin,
up to $\ord{1/m_B}$ corrections.

In this language, Stech's model amounts to considering only the
contribution from $\theta_2^{\pi,\mathcal{V}}$ and setting all other
$\theta$'s to zero.  In this sense, the model is a truncation of the
HQS results. As we shall see, its usefulness and justification rest on
the facts that it satisfies all known constraints and fits the lattice
data (available at large $q^2$) well.  Evaluating the traces in
\eqs{pseudo}{vector} and solving the resulting equations for the form
factors in a standard notation, we find
\bea
F_1(q^2) & = & \sqrt{\frac{m_B}{m_\pi}} \theta_2^\pi(v\cdot p_\pi)
\nonumber\\
F_0(q^2) & = & \l(1-\frac{q^2}{m_B^2-m^2_\pi}\r) F_1(q^2)
\nonumber\\
A_0(q^2) & = & \sqrt{\frac{m_B}{m_\mathcal{V}}} \theta_2^\mathcal{V}(v\cdot p_\mathcal{V})
\nonumber\\
V(q^2) & = & \l(1+\frac{m_\mathcal{V}}{m_B}\r)A_0(q^2)
\nonumber\\
A_1(q^2) & = & \frac{m_B^2+m_\mathcal{V}^2}{m_B(m_B+m_\mathcal{V})}\l(1-
\frac{q^2}{m_B^2+m_\mathcal{V}^2}\r) A_0(q^2)
\nonumber\\
A_2(q^2) & = & \frac{m_B+m_\mathcal{V}}{m_B}\l(1-\frac{2m_\mathcal{V}(m_B+m_\mathcal{V})}
{(m_B+m_\mathcal{V})^2-q^2}\r)A_0(q^2)
\nonumber\\
2T_1(q^2) & = & A_0(q^2)
\nonumber\\
2iT_2(q^2) & = & \l(1-\frac{q^2}{m_B^2-m_\mathcal{V}^2}\r) A_0(q^2)\ ,
\label{stech}
\eea
which are equivalent to the relations given in Eq.(11) of \re{stech1}
with, in addition, relations involving $T_1$ and $T_2$.

As we will see below, these relations can be used for a simple and
successful description of lattice results for the form factors. We
recall that, by construction, this model respects HQS scaling
relations and satisfies the two kinematic constraints:
\be
F_1(0) = F_0(0), \qquad T_1(0) = iT_2(0)
\ee
For the parametrization to be complete we must specify the functions
$\theta_2^{\pi,\mathcal{V}}$. This will be discussed in \sec{ffs}.

\section{Lattice Details}

The details of the lattice simulation can be found in \re{ukqcd:hlff}.
The chiral extrapolations for $\btorho$ and $\btokstargamma$ are
described in \re{ukqcd:btorho} while those for $\btopi$ are explained
in \re{lpl:btopi-bounds}.
The only way in which our determination of the lattice form factors
differs here is in the heavy-quark mass extrapolation. We improve this
extrapolation for the form factors $A_1$, $V$, $T_1$ and $T_2$ by
imposing heavy quark symmetry in the infinite heavy-quark mass limit,
as we now explain.

All form factors are calculated for four values of the heavy quark
mass around the charm mass and for a variety of values of $q^2$.  In
previous work \cite{ukqcd:hlff,ukqcd:btorho}, the form factors were
extrapolated at fixed four-velocity recoil, $\omega =
v\cdot(p_{\pi,\mathcal{V}}/m_{\pi,\mathcal{V}})$, near the zero recoil
point $\omega=1$, by fitting to the following heavy-quark scaling
relations:
\be
f\Theta M^{N_f/2}=\gamma_f(1+\frac{\delta_f}{M}+\frac{\epsilon_f}{M^2}+
\cdots)
\label{eq:extrap}
\ee
where $f=F_1,F_0,A_0,V,A_1,A_2,T_1,T_2$ and $N_f=-1,1,-1,-1,1,-1,-1,1$
respectively.  Here, $M$ is the mass of the heavy-light meson,
$\gamma_f$, $\delta_f$ and $\epsilon_f$ are the fit
parameters\footnote{$\epsilon_f$ is set to 0 for linear
extrapolations.}, and $\Theta$ comes from leading logarithmic matching
and is chosen to be 1 at the $B$ mass~\cite{neubert:physrep},
\be
\label{eq:theta}
\Theta = \Theta(M/m_B) = \left( \frac{\alpha_s(M)}{\alpha_s(m_B)} 
                         \right)^{\frac{2}{\beta_0}},
\ee
with $\beta_0 = 11$ in the
quenched approximation and $\Lambda_{\rm QCD} = 200\mev$. 

While correct, this procedure neglects the fact that in the $M\to\infty$
limit, HQS predicts
\be
A_1(q^2)=2iT_2(q^2),\quad V(q^2)=2T_1(q^2)
\label{eq:a1t2vt1hqs}
\ee
for $q^2$ not too far from $\qsqmax$, i.e.\ close to zero recoil.  In
the present paper, we enforce this prediction of HQS by performing a
combined fit, at fixed $\omega$, of the pairs of form factors
$(A_1,T_2)$ and $(V,T_1)$ to the parametrizations of \eq{eq:extrap}
with the following constraints on the fit parameters:
\be
\gamma_{A_1}=2i\gamma_{T_2},\quad \gamma_V=2\gamma_{T_1}
\ .
\ee
This not only guarantees that the extrapolated form
factors are consistent with HQS in the infinite mass limit, but also
reduces statistical errors, because the number of parameters is
decreased.

\begin{figure}
\unit=0.8\hsize
\hbox to\hsize{\hfill
\vbox{\offinterlineskip
\epsfxsize=\unit\epsffile{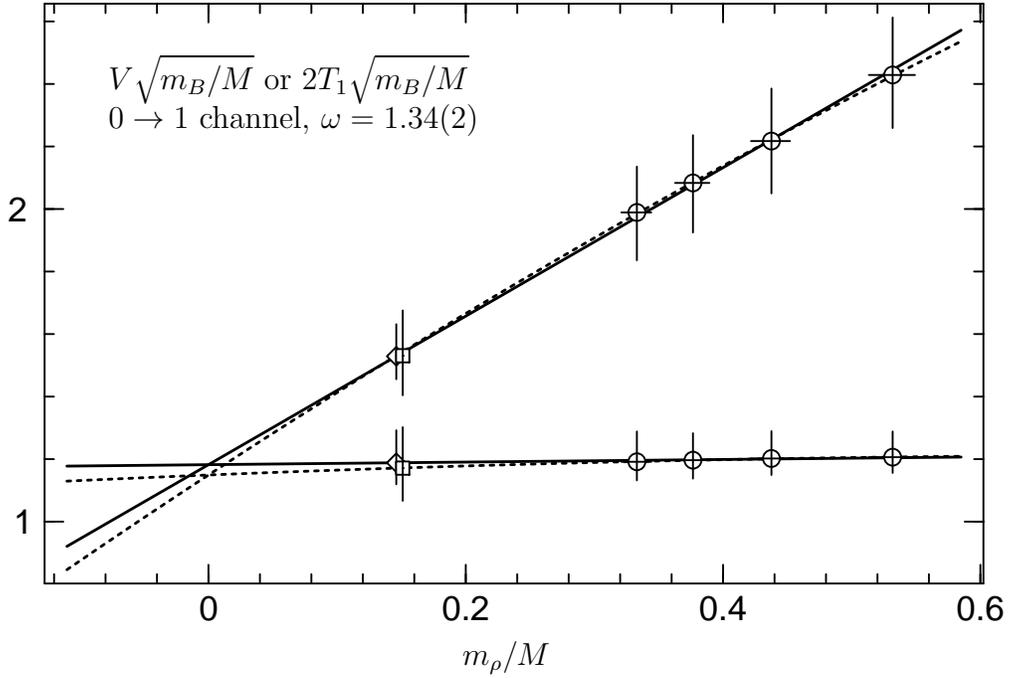}
\point 0.1 0.575 {V \sqrt{m_B/M} \mathrm{\ or\ } 2T_1\sqrt{m_B/M}}
\point 0.1 0.525 {0\to1 \mathrm{\ channel,\ }\omega = 1.34(2)}
\kern0.5em
\hbox to\unit{\hfill$m_\rho/M$\hfill}
}\hfill}
\caption[]{Constrained $1/M$ extrapolation of the pair of form factors
$(V,2T_1)$, as described in the text. The solid curves correspond to a
linear extrapolation, the dashed curves to a quadratic fit.  The four
lower circle points are the lattice results for $2T_1 \sqrt{m_B/M}$
and the four upper ones are those for $V \sqrt{m_B/M}$, in the case
where a $B$ meson at rest decays to produce a $\rho$ meson with one
(lattice) unit of three momentum. The points at $m_\rho/M=0.146$
correspond to the linearly (diamonds) and quadratically (squares)
extrapolated results for the two form factors at $M=m_B$.
\label{fig:vt1-2}}
\end{figure}

As an example, we show in \fig{fig:vt1-2} the combined fit of
$(V,2T_1)$ to the constrained heavy quark scaling relation just
described, for the case of a final state $\rho$. Both the linear and
quadratic fits in $1/M$ are excellent, confirming our earlier finding,
in \re{ukqcd:btorho}, that the form factors $A_1$, $V$, $T_1$ and
$T_2$ satisfy the infinite mass relations of \eq{eq:a1t2vt1hqs} very
well, even when extrapolated independently.  Because linear and
quadratic extrapolations always give results which agree within error,
because the quadratic term is always consistent with 0 and because the
quadratic fits may emphasize small discretization errors, we will use
the linear results in the following.

In the results reported below, all errors are statistical central 68\%
bounds. For decays to a light final state vector we use 250 bootstrap
samples from our lattice data. For decays to a light pseudoscalar we
add errors to account for the chiral
extrapolation~\cite{lpl:btopi-bounds} and propagate them assuming
standard Gaussian statistics.

\section{Form Factors}\label{ffs}

One could imagine performing a combined fit to $\btopi$, $\btorho$ and
$\btokstargamma$ form factors assuming
$\theta_2^\pi(q^2){=}\theta_2^\mathcal{V}(q^2)$ in \eq{stech}.
However, there is no reason to have spin symmetry relating the final
pseudoscalar and vector states.  Furthermore, our lattice results
clearly indicate that the $q^2$ dependences of $F_1(q^2)$ and
$A_0(q^2)$ are different, in contradiction to the assumption
$\theta_2^\pi(q^2) = \theta_2^\mathcal{V}(q^2)$. Thus we consider
decays to pseudoscalar and vector states separately.

\subsection{$\btorho$ and $\btokstargamma$ decays}

So as {\it not} to assume flavour $SU(3)$ symmetry in our combined
description of $\btorho$ and $\btokstargamma$ decays we have used the
freedom to adjust quark masses in lattice calculations and considered
two situations:
\begin{itemize}
\item[{\bf A}] The mass of the light active quark $q$ in \eq{vector}
is set to the $u$ quark mass (i.e. $\mathcal{V}=\rho$ in
\eqs{vector}{stech}). In this case, the combined fit to the form
factors of the operators $\bar q \gamma^\mu(1{-}\gamma^5) b$ and $\bar
q q_\nu\sigma^{\mu\nu}(1{+}\gamma^5)b/2$ constrains, through \eq{stech},
the form factors relevant for the decay $\btorho$.
\item[{\bf B}] The mass of the light active quark $q$ in \eq{vector}
is set to the $s$ quark mass (i.e. $\mathcal{V}=K^*$ in
\eqs{vector}{stech}). Now the combined fit constrains, through
\eq{stech}, the form factors relevant for the decay $\btokstargamma$.
\end{itemize}

\begin{figure}
\unit=0.8\hsize
\hbox to\hsize{\hfill
\vbox{\offinterlineskip
\epsfxsize=\unit\epsffile{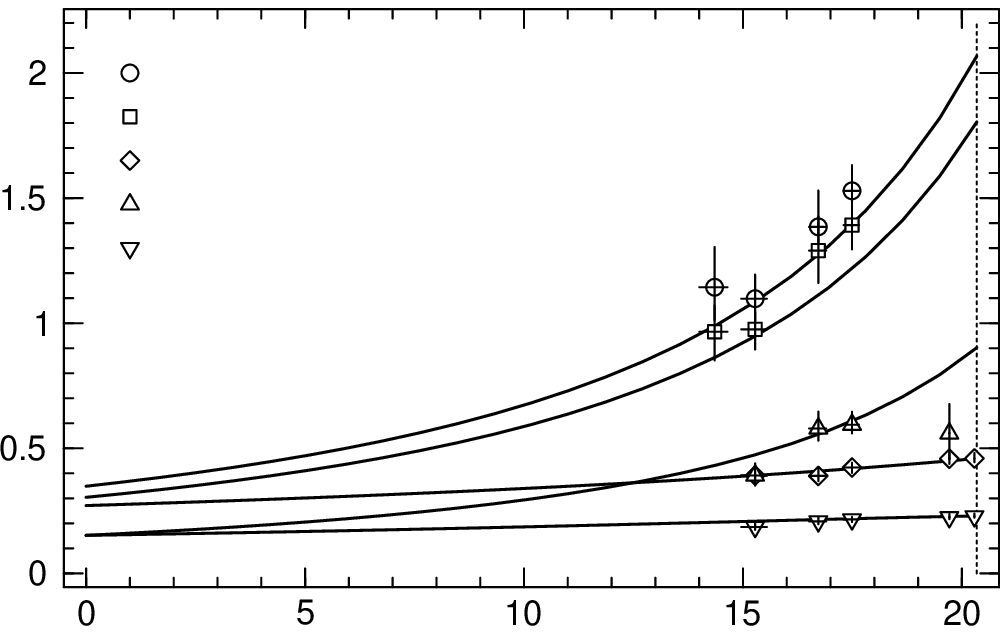}
\point 0.155 0.58 V
\point 0.155 0.536 {A_0}
\point 0.155 0.492 {A_1}
\point 0.155 0.448 {T_1}
\point 0.155 0.404 {T_2}
\hbox to\unit{\hfill$q^2 \; (\gev^2)$\hfill}
}\hfill}
\caption[]{Fit to the lattice predictions for $A_0$, $A_1$, $V$, $T_1$
and $T_2$ for a $\rho$ meson final state (Situation A) using the
parametrization of \protect\eq{stech} and assuming a pole form for
$A_1$ as given in \eq{a1pole}. The dashed vertical line indicates
$\qsqmax$.
\label{fig:btorho}}
\end{figure}

For the parametrization of \eq{stech} to be complete we must specify
the function $\theta_2^{\mathcal{V}}$ or, equivalently, one of the
form factors. With pole dominance ideas in mind, we shall consider a
pole form for $A_1(q^2)$ which is compatible with previous lattice
analyses~\cite{ukqcd:btorho} and LCSR results~\cite{ballbraun:lcsr}:
\be
A_1(q^2) = 
\frac{A_1(0)}{1-q^2/M_1^2}
\label{a1pole}\ ,\ee
where the two free parameters are $A_1(0)$ and $M_1$, a mass on the
order of $m_B$. With these two parameters alone, we can describe the
six form factors needed for decays to a vector final state\footnote{In
practice, we fit lattice data for the five form factors $V$, $A_0$,
$A_1$, $T_1$ and $T_2$ to fix our free parameters. Our present lattice
results do not allow a reliable extraction of $A_2$.}. This form
further guarantees that all form factors scale like $m_B^{-3/2}$ at
$q^2=0$ as predicted by LCSR in~\re{ballbraun:lcsr}.  The results for
$\btorho$ and $\btokstargamma$ decays are presented in
\tab{tab:btorho-and-btokstargamma}, and \figs{fig:btorho}{fig:bsg}. As
the $\chi^2$ indicates, this simple parametrization works surprisingly
well. The $SU(3)$ flavour dependence in our results, from comparing
situations A and B, is mild. Corresponding form factor values at
$q^2=0$ differ by less than 10\%.

\begin{table}
\tabcolsep0.93\tabcolsep
\hbox to\hsize{\hfill
\begin{tabular}{lcccc}
\multicolumn{5}{c}{$\btorho$}\\
\hline
\multicolumn{5}{c}{$A_1(0) = 0.27\err{0.05}{0.04}$}\\
\multicolumn{5}{c}{$M_1 = 7.0\err{1.2}{0.6}\gev$}\\
\multicolumn{5}{c}{$\chi^2/\dof = 24/20$}\\
\hline
& $A_1$ & $A_2$ & $A_0$ & $V$\\ \hline
$q^2{=}0$ & $0.27\err{0.05}{0.04}$ & $0.26\err{0.05}{0.03}$  &
 $0.30\err{0.06}{0.04}$ & $0.35\err{0.06}{0.05}$\\
$\qsqmax$ & $0.46\err{0.02}{0.01} $ & $0.88\err{0.05}{0.03}$ & 
$1.80\err{0.09}{0.05}$ & $2.07\err{0.11}{0.06}$
\\ \hline
\end{tabular}
\hfill
\begin{tabular}{lcc}
\multicolumn{3}{c}{$\btokstargamma$}\\
\hline
\multicolumn{3}{c}{$A_1(0) = 0.29\err{0.04}{0.03}$}\\
\multicolumn{3}{c}{$M_1 = 6.8\err{0.7}{0.4}\gev$}\\
\multicolumn{3}{c}{$\chi^2/\dof = 27/20$}\\
\hline
& $T_1$ & $T_2$ \\ \hline
$q^2{=}0$ & $0.16\err{0.02}{0.01}$ & \\
$\qsqmax$ & $0.90\err{0.05}{0.04}$ & $0.25\err{0.01}{0.01}$\\ \hline
\end{tabular}
\hfill}
\caption[]{Results of fits to the lattice predictions for $A_0$,
$A_1$, $V$, $T_1$ and $T_2$ assuming a pole form for $A_1$, as given
in \eq{a1pole}, in the parametrization of \protect\eq{stech}. On the
left are results for a $\rho$ meson final state (Situation A) and on
the right, results for a $K^*$ meson final state (Situation B). We
also provide the values of the form factors for $\btorho$ and
$\btokstargamma$ decays at $q^2{=}0$ and $\qsqmax$, as determined by
the fit.
\label{tab:btorho-and-btokstargamma}}
\end{table}

\begin{figure}
\unit=0.8\hsize
\hbox to\hsize{\hfill
\vbox{\offinterlineskip
\epsfxsize=\unit\epsffile{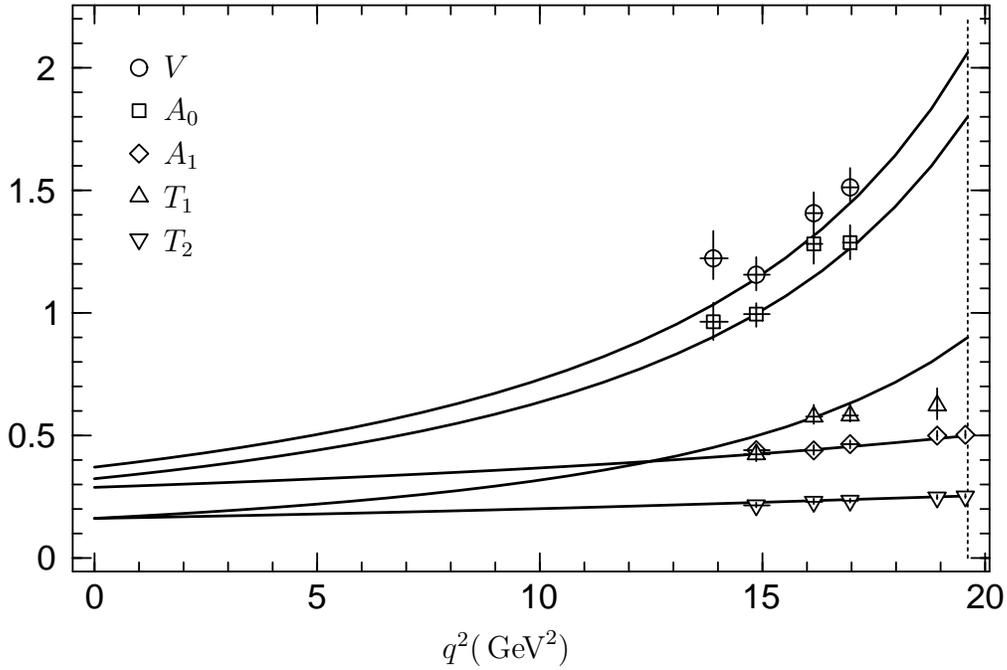}
\point 0.155 0.58 V
\point 0.155 0.536 {A_0}
\point 0.155 0.492 {A_1}
\point 0.155 0.448 {T_1}
\point 0.155 0.404 {T_2}
\hbox to\unit{\hfill$q^2 (\gev^2)$\hfill}
}\hfill}
\caption[]{Fit to the lattice predictions for $A_0$, $A_1$, $V$, $T_1$
and $T_2$ for a $K^*$ meson final state (Situation B) using
the parametrization of \protect\eq{stech} and assuming a pole form for
$A_1$ as given in \eq{a1pole}. The dashed vertical line indicates
$\qsqmax$.
\label{fig:bsg}}
\end{figure}

We have considered functional dependences for $A_1$ different from
that in \eq{a1pole}. Constant behaviour has already been ruled out by
lattice results~\re{ukqcd:btorho}.  Dipole (and in general higher powers:
tripole, \ldots) and pole fits are hardly distinguishable in the
physical range $0 \leq q^2 \leq \qsqmax$. In a dipole fit, the mass
parameter $M_{\rm dipole}$ is roughly given by $M_{\rm dipole} \approx
\sqrt2 M_{\rm pole}$ and hence pole and dipole fits agree in the range
of values of $q^2$ explored in \figs{fig:btorho}{fig:bsg}.
We have also studied ``modified-constant'' behaviour for $A_1$:
\be
A_1(q^2)=(1-q^2/m_\alpha^2) \frac{A_1(0)}{1-q^2/m_\beta^2}
\ ,
\label{modcst}
\ee
where $m_\alpha$ and $m_\beta$ are masses on the order of $m_B$. Such
a behaviour was introduced in \re{lpl:btopi-bounds} in the context of
$B\to\pi l\nu$ decays. Within the framework of \eq{stech}, this can
be implemented by parametrizing $A_0(q^2)$ as a pole without
increasing the number of free parameters.  Thus, we have performed a
second set of fits where we complete the parametrization of \eq{stech}
by taking the following functional form for $A_0$:
\be
A_0(q^2) = 
\frac{A_0(0)}{1-q^2/M_0^2}
\label{a0pole}
\ee
where the two free parameters are $A_0(0)$ and $M_0$, a mass of the
order of $m_B$.  The corresponding results for $\btorho$ and
$\btokstargamma$ decays are presented in
\tab{tab:btorho-and-btokstargamma-2}.  This parametrization yields
different behaviour for the form factors at small $q^2$: the values at
$q^2=0$ differ by two to three standard deviations.

\begin{table}
\tabcolsep0.93\tabcolsep
\hbox to\hsize{\hfill
\begin{tabular}{lcccc}
\multicolumn{5}{c}{$\btorho$}\\
\hline
\multicolumn{5}{c}{$A_0(0) = 0.40\err{0.04}{0.03}$}\\
\multicolumn{5}{c}{$M_0 = 5.10\err{0.08}{0.06}\gev$}\\
\multicolumn{5}{c}{$\chi^2/\dof = 25/20$}\\
\hline
& $A_1$ & $A_2$ & $A_0$ & $V$\\ \hline
$q^2{=}0$ & $0.35\err{0.03}{0.02}$ & $0.34\err{0.03}{0.02}$ &
 $0.40\err{0.04}{0.03}$ & $0.45\err{0.04}{0.03}$ \\
$\qsqmax$ & $0.46\err{0.02}{0.01}$ & $0.89\err{0.05}{0.03}$ 
& $1.81\err{0.10}{0.06}$ & $2.08\err{0.11}{0.06}$
\\ \hline
\end{tabular}
\hfill
\begin{tabular}{lcc}
\multicolumn{3}{c}{$\btokstargamma$}\\
\hline
\multicolumn{3}{c}{$A_0(0) = 0.42\err{0.03}{0.02}$}\\
\multicolumn{3}{c}{$M_0 = 5.05\err{0.06}{0.05}\gev$}\\
\multicolumn{3}{c}{$\chi^2/\dof = 27/20$}\\
\hline
& $T_1$ & $T_2$ \\ \hline
$q^2{=}0$ & $0.21\err{0.01}{0.01}$ & \\
$\qsqmax$ & $0.91\err{0.05}{0.04}$ & $0.25\err{0.01}{0.01}$\\ \hline
\end{tabular}\hfill}
\caption[]{Results of fits to the lattice predictions for $A_0$,
$A_1$, $V$, $T_1$ and $T_2$ assuming a pole form for $A_0$, as given
in \eq{a0pole}, in the parametrization of \protect\eq{stech}. On the
left are results for a $\rho$ meson final state (Situation A) and on
the right, results for a $K^*$ meson final state (Situation B). We
also provide the values of the form factors relevant for $\btorho$ and
$\btokstargamma$ decays at $q^2{=}0$ and $\qsqmax$ as determined by
the fit.
\label{tab:btorho-and-btokstargamma-2}}
\end{table}

Though we cannot discriminate against this parametrization on the
basis of $\chi^2$ alone, there are physical arguments for preferring
the $A_1$ pole fits.  Fixing pole behaviour for $A_0$ in the context
of Stech's parametrization forces all form factors to diverge at the
same value of $q^2$ even though different form factors receive
contributions from resonances with different quantum numbers and
masses.  Furthermore, we find that this divergence occurs at around
$5.1\gev$, below the first physical pole, the $B$.  Assuming pole
behaviour for $A_1$, on the other hand, allows the form factors which
receive contributions from $1^+$ resonances ($A_1$, $A_2$ and $T_2$)
to diverge at larger values of $q^2$ than the more singular form
factors $V$, $T_1$ and $A_0$. Our fit confirms such behaviour as we
find $M_1=7.0\err{1.2}{0.6} \gev$, in reasonable agreement with
lattice~\cite{ape:hl-semilept} and quark model
estimates~\cite{ehq:hl-mesons,ehq2:hl-mesons,ehq3:hl-mesons} for the
$1^+$, $b\bar q$ resonance.  The more singular form factors then
diverge at $q^2=m_B^2+m_\mathcal{V}^2$ which is close to the physical
$0^-$ ($A_0$) and $1^-$ ($V$ and $T_1$) poles.  Finally, LCSR scaling
relations at $q^2=0$ combined with HQS requirements at $\qsqmax$ rule
out pole behaviour for $A_0$ but not for $A_1$. Thus, for physical
applications, we consider only the parametrization given in
\eqs{stech}{a1pole}.

\subsection{$\btopi$ decays}

Stech's parametrization predicts that $F_0(\qsqmax)$ vanishes in the
chiral limit, in contradiction with our results and made unlikely by
unitarity bounds~\cite{lpl:btopi-bounds}. Furthermore, the $B^*$ which
contributes a pole to $F_1$ induces the same singularity in $F_0$
within Stech's model.  Because the $B^*$ pole is very close to
$\qsqmax$, it provokes a much more pronounced $q^2$ dependence for
$F_0$ than observed in the lattice results or induced by the nearest
resonance in this $0^+$ channel whose mass is significantly larger
than $m_{B^*}$.

Restricting to polar-type $q^2$-dependences, consistent with the
kinematical constraint, $F_1(0)=F_0(0)$, heavy quark symmetry and
unitarity bounds~\cite{lpl:btopi-bounds,ukqcd:hlff}, we consider two
possible functional forms:
\begin{itemize}
\item {\bf pole/dipole} 
\be
F_1(q^2) = \frac{F(0)}{\l(1-q^2/m_1^2\r)^{2}}\, , \qquad
F_0(q^2) = \frac{F(0)}{\l(1-q^2/m_0^2\r)} \label{f1dipole}
\ee
\item{\bf ``modified-constant''/pole}
\be
F_1(q^2) = \frac{F(0)}{\l(1-q^2/m_1^2\r)}\, , \qquad
F_0(q^2) =  F(0) \frac{(1-q^2/m_2^2)} {\l(1-q^2/m_0^2\r)}
\ee
\end{itemize}
These two dependences have been studied previously in \re{ukqcd:hlff}
with light quark masses slightly larger than that of the strange quark
and in \re{lpl:btopi-bounds} for massless $u$ and $d$ quarks.  The fit
results are summarized here in \tab{tab:btopires}\footnote{The errors
quoted here differ from those in~\cite{lpl:btopi-bounds} which were
determined by an incorrect procedure. The central fit values agree.
This difference does not affect the dispersive bounds of
\re{lpl:btopi-bounds}, nor the comparison of these bounds with the
various parametrizations.}.  We note that only the pole-dipole
parametrization is consistent with LCSR scaling relations at $q^2=0$
together with HQS requirements at $\qsqmax$. Both fits are consistent
with the dispersive bounds of \re{lpl:btopi-bounds}.

\begin{table}[tb]
\begin{center}
\begin{tabular}{lccccc}
\hline 
fit type & $F(0)$ &  
$m_1\,(\gev)$ & $m_0\,(\gev)$ & $m_2\,(\gev)$ & $\chi^2/\dof$\\ \hline
``modified-const''  & $0.43\pm 0.06$ 
& $m_{B^*}$ & 5.5 & $5.89\pm 0.52$  & 0.5/4\\ 
pole/dipole & $0.27\pm 0.11$ & $5.79\pm 0.58$  & $6.1\pm 1.5$ & --
& 0.1/3 \\ \hline
\end{tabular}
\caption[]{Fits of the lattice results for $F_1(q^2)$ and $F_0(q^2)$ to
the parametrizations described in the text (taken from
\protect\re{lpl:btopi-bounds}).  In order to perform the
``modified-constant/pole'' fit with a limited range of results, $m_1$
and $m_0$ were fixed to the value of the corresponding nearest
resonances. (This latter fit 
is named ``fixed-pole'' in \protect\re{lpl:btopi-bounds}.)
\label{tab:btopires}}
\end{center}
\end{table}

\section{Phenomenological Consequences}

\begin{figure}
\unit=0.95\hsize
\hbox to\hsize{\hfill
\vbox{\offinterlineskip
\epsfxsize=\unit\epsffile{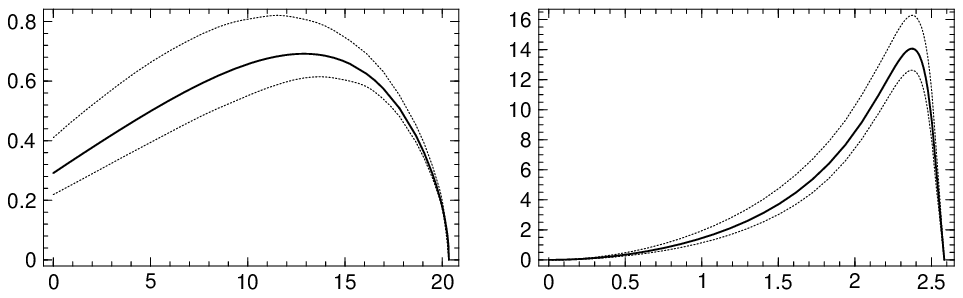}
\point 0.06 0.08 {d\Gamma/dq^2 \; (\vub^2 10^{-12}\gev^{-1})}
\point 0.6 0.283 {d\Gamma/dE \; (\vub^2 10^{-12})}
\kern0.5em
\hbox to\unit{\hbox to0.5\unit{\hfill$q^2 (\gev^2)$\hfill}
\hfill\hbox to0.45\unit{\hfill$E (\gev)$\hfill}}
}\hfill}
\caption[]{Differential decay spectra for $\btorho$ for massless
leptons: (a) $d\Gamma/dq^2$ in units of $10^{-12} \vub^2 \gev^{-1}$,
(b) $d\Gamma/dE$ in units of $10^{-12} \vub^2$. The form factors are
taken from the fit to the parametrization of \protect\eq{stech},
assuming a pole form for $A_1$ as given in \eq{a1pole}. The dashed
lines show the envelope of the 68\% bootstrap errors computed
separately for each value of $q^2$ or $E$
respectively.\label{fig:btorho-spectra}}
\end{figure}

\begin{figure}
\unit=0.95\hsize
\hbox to\hsize{\hfill
\vbox{\offinterlineskip
\epsfxsize=\unit\epsffile{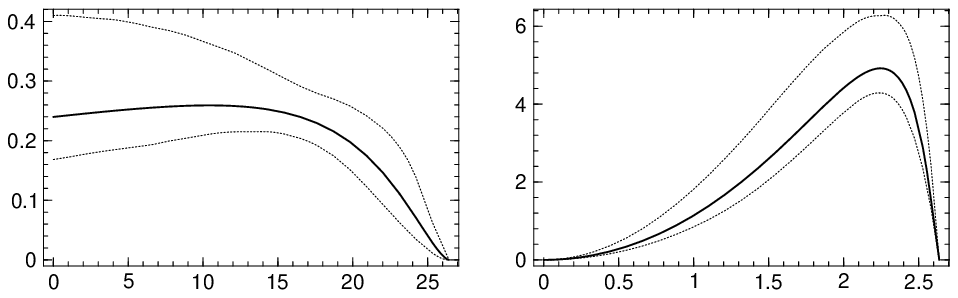}
\point 0.06 0.08 {d\Gamma/dq^2 \; (\vub^2 10^{-12}\gev^{-1})}
\point 0.6 0.283 {d\Gamma/dE \; (\vub^2 10^{-12})}
\kern0.5em
\hbox to\unit{\hbox to0.5\unit{\hfill$q^2 (\gev^2)$\hfill}
\hfill\hbox to0.45\unit{\hfill$E (\gev)$\hfill}}
}\hfill}
\caption[]{Differential decay spectra for $\btopi$ for massless
leptons: (a) $d\Gamma/dq^2$ in units of $10^{-12} \vub^2 \gev^{-1}$,
(b) $d\Gamma/dE$ in units of $10^{-12} \vub^2$. The form factors are
taken from the fit with a dipole form for $F_1$ as given in
\eq{f1dipole}. The dashed lines show the envelope of the 68\%
errors computed separately for each value of $q^2$ or $E$
respectively.
\label{fig:btopi-spectra}}
\end{figure}

Using the preferred pole form for $A_1$ in the decay $\btorho$, we
find the differential decay rate spectra in $q^2$ and the lepton
energy $E$.\footnote{See \re{ks} for the decay rate kinematics,
including the case of non-zero lepton masses.} These are shown in
\fig{fig:btorho-spectra}. Likewise, using the pole/dipole fits for
$\btopi$, we find the differential spectra shown in
\fig{fig:btopi-spectra}. Integrating to find the total decay rates we
have, for massless leptons in the final state:
\begin{eqnarray}
\Gamma(\btorho)/\vub^2 &=& 10.9 \err{2.3}{1.5} \times 10^{-12} \gev =
 16.5 \err{3.5}{2.3} \,\mathrm{ps}^{-1} \\
\Gamma(\btopi)/\vub^2 &=& 5.6\err{2.2}{0.6}  \times 10^{-12} \gev =
 8.5\err{3.4}{0.9} \,\mathrm{ps}^{-1}
\end{eqnarray}
For decays with a tau lepton in the final state we find:
\begin{eqnarray}
\Gamma(\bar B^0\to\rho^+ \tau^- \bar\nu_\tau)/\vub^2 &=&
    5.8 \err{0.9}{0.6} \times 10^{-12} \gev =
    8.8 \err{1.4}{0.9} \,\mathrm{ps}^{-1} \\
\Gamma(\bar B^0\to\pi^+ \tau^- \bar\nu_\tau)/\vub^2 &=&
    3.8 \err{1.2}{0.2}  \times 10^{-12} \gev =
    5.8 \err{1.8}{0.4} \,\mathrm{ps}^{-1}
\end{eqnarray}

In \tab{tab:compare} we compare our results with recent LCSR
\cite{ballbraun:lcsr,alibraunsimma:lcsr,bkr:lcsr,bbkr:lcsr,%
kr-ichep96:lcsr,becirevic}
calculations\footnote{We quote LCSR results at leading order in
perturbative QCD. The $\ord{\alpha_s}$ corrections to the leading
twist term only are now known for $F_1$~\cite{krwy:lcsr}.}.  The
agreement of the form factor values at $q^2=0$ is very good. In the
spectrum $d\Gamma(\btorho)/dq^2$ we note that our result is larger
near $q^2=0$, as is our value for the ratio $\Gamma_L/ \Gamma_T$. The
differential decay rate at $q^2=0$ is proportional to
\be A_0(0) = {m_B + m_\rho\over 2 m_\rho}
A_1(0) - {m_B - m_\rho\over 2 m_\rho} A_2(0) 
\ ,\ee
which is the longitudinal contribution, so that small differences in
$A_1(0)$ and $A_2(0)$ can lead to significant differences in the
differential rate and in the ratio of longitudinal to transverse
rates. Since the lattice results are available for large $q^2$, the
$d\Gamma/dq^2$ plots in \figs{fig:btorho-spectra}{fig:btopi-spectra}
also show errors growing larger towards $q^2=0$. This is complementary
behaviour to LCSR results, which are more reliable at low $q^2$.

All other rates agree with the LCSR results within errors. Agreement
is also excellent with the experimental results of
CLEO~\cite{cleo:semilept96}. For the ratio $\Gamma(\rho)/\Gamma(\pi)$,
which is independent of $|V_{ub}|$ and can therefore be compared
directly to the value in \tab{tab:compare}, CLEO quotes
$1.4\err{0.6}{0.4}\pm 0.3\pm 0.4$, where the errors are statistical,
systematic and the estimated model dependence. Moreover, branching
ratios, obtained from the rates in \tab{tab:compare} with the values
of $|V_{ub}|$ and of the $B$ lifetimes given in \re{cleo:semilept96}
by CLEO, compare very favorably with the values in that same
publication.

Finally, the additional constraints provided by our approach may shed
some light on the twofold ambiguity in the lattice value of $T_1(0)$
(see for example \cite{ukqcd:hlff,zako94}) by favouring the smaller of
the two determinations.  With our prediction for $T_1(0)$, we can
determine the hadronization ratio $R_{K^*}=\Gamma(B\to
K^*\gamma)/\Gamma(b\to s\gamma)$, which is given, up to
$\ord{1/m_b^2}$ corrections, by \cite{ciuchini94}
\be
R_{K^*} = 4\l(\frac{m_B}{m_b}\r)^3\l(1-\frac{m^2_{K^*}}{m^2_B}\r)^3
|T_1(0)|^2
\ .\ee
We find $R_{K^*}=(16\err{4}{3})\%$, to be compared with the experimental
value $(18\pm 7)\%$ \cite{CLEO:ammar96}.

\begin{table}
\begin{center}
\hbox to\hsize{\hss
\begin{tabular}{l>{$}l<{$}>{$}l<{$}>{$}l<{$}>{$}l<{$}>{$}l<{$}>{$}l<{$}}
\hline
 & F_1(0) & A_1(0) & A_2(0) & A_0(0) & V(0) & T_1(0)\\
\hline
This work & 0.27\pm0.11 & 0.27\err{0.05}{0.04} & 0.26\err{0.05}{0.03}
 & 0.30\err{0.06}{0.04} & 0.35\err{0.06}{0.05} & 0.16\err{0.02}{0.01}\\
LCSR \cite{ballbraun:lcsr} &
 & 0.27\pm0.05 & 0.28\pm0.05 & & 0.35\pm0.07 \\
LCSR \cite{alibraunsimma:lcsr} &
 & 0.24\pm0.04 & & & 0.28\pm0.06 & 0.16\pm0.03 \\
LCSR \cite{bkr:lcsr,bbkr:lcsr} & 0.24$--$0.29 \\
LCSR \cite{becirevic} & & & & & & 0.15 \pm 0.03 \\
\hline
\end{tabular}\hss}
\end{center}
\begin{center}
\begin{tabular}{l>{$}l<{$}>{$}l<{$}>{$}l<{$}>{$}l<{$}}
\hline
 & \Gamma(\btopi) & \Gamma(\btorho) & \Gamma(\rho)/\Gamma(\pi) &
 \Gamma_L/\Gamma_T\\
\hline
This work & 8.5\err{3.3}{1.4} & 16.5\err{3.5}{2.3} & 1.9\err{0.9}{0.7} & 
0.80\err{0.04}{0.03} \\
LCSR \cite{ballbraun:lcsr} &
  & 13.5\pm4.0 & 1.7\pm0.5 & 0.52\pm0.08\\
LCSR \cite{kr-ichep96:lcsr} &
  8.7 \\
\hline
\end{tabular}
\end{center}
\caption[]{Form factor values at $q^2 =0$ and decay rates and ratios
for $b\to u$ transitions from this calculation and from light cone sum
rules (LCSR). Decay rates are given in units of $\vub^2
\,\mathrm{ps}^{-1}$.}
\label{tab:compare}
\end{table}

\section{Conclusion}

Inspired by the work of Stech~\cite{stech1}, we have designed a simple
parametrization for the form factors which describe $B\to V$
transitions, where $V$ is a light vector meson.  This parametrization
is consistent with heavy quark symmetry and kinematical constraints
but requires an ansatz for the $q^2$-dependence of one of the form
factors. The parameters of the ansatz are determined by fitting to
lattice results around $\qsqmax$. We have explored several ans\"atze
and favour pole behaviour for $A_1$: though not singled out by
$\chi^2$ alone, it satisfies LCSR scaling relations at $q^2=0$ and
allows form factors that receive contributions from resonances with
different quantum numbers and masses to diverge at different $q^2$.
As a result we describe semileptonic and radiative $B$ decays to the
same light vector meson with only two parameters.  We use the freedom
provided by lattice calculations to consider two independent
situations. In the first, the mass of the final state vector meson is
set to the physical $\rho$ mass.  We then apply our model and fit both
the semileptonic and radiative form factors simultaneously, keeping
the former to describe the physical $\btorho$ decay.  In the second
situation, the final state meson is chosen to be a $K^*$.  Again we
fit all form factors at once, this time keeping the radiative form
factor results to describe the physical $\btokstargamma$ decay.  In
this way, we do not assume $SU(3)$ flavour symmetry.

For semileptonic $B$ transitions to a light pseudoscalar meson, we
perform a separate analysis, since we do not assume spin symmetry for
the final state meson. Our preferred description, consistent with
heavy quark symmetry, kinematic constraints and LCSR scaling
relations, is a pole/dipole fit with three parameters. We apply this
to $\btopi$ decays.

Even though our approach requires some assumptions, we have tried to
minimize their number and make them consistent with all known
theoretical constraints and lattice results.  The resulting
parametrizations should be extremely useful for phenomenological
applications because they provide a simple and effective description
of the $q^2$ behaviour of the various form factors.

We should also like to stress that we have introduced an improved
procedure for extrapolating form factors in heavy-quark mass from
around the charm, where they are calculated on the lattice, to
the $b$-quark mass.  This procedure reduces statistical errors
by making full use of the constraints
of heavy quark symmetry.

\section*{Acknowledgements}

We thank other members of the UKQCD collaboration for the original
calculations of the lattice correlation functions.
%
We acknowledge the Particle Physics and Astronomy Research Council
(PPARC) for travel support under grant GR/L29927.
JMF is supported by PPARC under grant GR/K55738 and thanks the British
Council for travel support under Acciones Integradas grant 3241.
LL thanks the Minist\`ere des Affaires Etrang\`eres for travel
support under grant PAI-Picasso $N^o$97088. JN acknowledges support
from DGES contract PB95-1204 and Acciones Integradas contracts
HF1996-0155 and HB1996-0001.

\bibliographystyle{elsevier}
\bibliography{stech}

\begin{thebibliography}{10}

\bibitem{cleo:semilept96}
{CLEO} collaboration, J.P. Alexander et~al.,
\newblock Phys. Rev. Lett. 77 (1996) 5000.

\bibitem{mad1}
M.A. Diaz,
\newblock Phys. Lett.~B 304 (1993) 278, hep-ph/9303280.

\bibitem{mad2}
M.A. Diaz,
\newblock Phys. Lett.~B 322 (1994) 207, hep-ph/9311228.

\bibitem{hewett:topten}
J.L. Hewett,
\newblock SLAC preprint SLAC--PUB--6521 (1994), hep-ph/9406302.

\bibitem{htt:ipnp}
J.L. Hewett, T. Takeuchi and S. Thomas,
\newblock SLAC preprint SLAC--PUB--7088 (1996), hep-ph/9603391.

\bibitem{stech1}
B. Stech,
\newblock Phys. Lett.~B 354 (1995) 447, hep-ph/9502378.

\bibitem{lpl:btopi-bounds}
L. Lellouch,
\newblock Nucl. Phys.~B 479 (1996) 353, hep-ph/9509358.

\bibitem{ukqcd:hlff}
{UKQCD} collaboration, D.R. Burford et~al.,
\newblock Nucl. Phys.~B 447 (1995) 425, hep-lat/9503002.

\bibitem{ukqcd:btorho}
{UKQCD} collaboration, J.M. Flynn et~al.,
\newblock Nucl. Phys.~B 461 (1996) 327, hep-ph/9506398.

\bibitem{ukqcd:a0a3}
{UKQCD} collaboration, J.M. Flynn and J. Nieves,
\newblock Nucl. Phys.~B 476 (1996) 313, hep-ph/9602201.

\bibitem{iw:hqet}
N. Isgur and M.B. Wise,
\newblock Phys. Rev.~D 42 (1990) 2388.

\bibitem{gmm}
P.A. Griffin, M. Masip and M. Mc{G}uigan,
\newblock Phys. Rev.~D 42 (1994) 5751, hep-ph/9312262.

\bibitem{neubert:physrep}
M. Neubert,
\newblock Phys. Rep. 245 (1994) 259, hep-ph/9306320.

\bibitem{ballbraun:lcsr}
P. Ball and V.M. Braun,
\newblock Phys. Rev.~D 55 (1997) 5561, hep-ph/9701238.

\bibitem{ape:hl-semilept}
{APE} collaboration, C.R. Allton et~al.,
\newblock Phys. Lett.~B 345 (1995) 513, hep-lat/9411011.

\bibitem{ehq:hl-mesons}
E.J. Eichten, C.T. Hill and C. Quigg,
\newblock Phys. Rev. Lett. 71 (1993) 4116, hep-ph/9308337.

\bibitem{ehq2:hl-mesons}
E.J. Eichten, C.T. Hill and C. Quigg,
\newblock Fermilab preprint FERMILAB-CONF-94-117-T (1994).

\bibitem{ehq3:hl-mesons}
E.J. Eichten, C.T. Hill and C. Quigg,
\newblock Fermilab preprint FERMILAB-CONF-94-118-T (1994).

\bibitem{ks}
J.G. K{\"o}rner and G.A. Schuler,
\newblock Phys. Lett.~B 231 (1989) 306.

\bibitem{alibraunsimma:lcsr}
A. Ali, V.M. Braun and H. Simma,
\newblock Z. Phys.~C 63 (1994) 437, hep-ph/9401277.

\bibitem{bkr:lcsr}
V.M. Belyaev, A. Khodjamirian and R. R{\"u}ckl,
\newblock Z. Phys.~C 60 (1993) 349, hep-ph/9305348.

\bibitem{bbkr:lcsr}
V.M. Belyaev et~al.,
\newblock Phys. Rev.~D 51 (1995) 6177, hep-ph/9410280.

\bibitem{kr-ichep96:lcsr}
A. Khodjamirian and R. R{\"u}ckl,
\newblock Proc. {ICHEP 96}, 28th Int. Conf. on High Energy Physics, Warsaw,
  Poland, 25--31 July 1996, edited by Z. Ajduk and A.K. Wroblewski, pp.
  902--905, World Scientific, Singapore, 1997, hep-ph/9610367.

\bibitem{becirevic}
D. Becirevic,
\newblock LPTHE Orsay preprint LPTHE--Orsay 97/16 (1997), hep-ph/9707271.

\bibitem{krwy:lcsr}
A. Khodjamirian et~al.,
\newblock W{\"u}rzburg, MPI M{\"u}nchen and Saclay preprint WUE/ITP--97--015,
  MPI--PhT/97--34, SPhT--T97/042 (1997), hep-ph/9706303.

\bibitem{zako94}
L. Lellouch,
\newblock Acta Phys. Polon. 25 (1994) 1679, hep-ph/9412284.

\bibitem{ciuchini94}
M. Ciuchini et~al.,
\newblock Phys. Lett.~B 334 (1994) 137, hep-ph/9401277.

\bibitem{CLEO:ammar96}
{CLEO} collaboration, R. Ammar et~al.,
\newblock CLEO preprint CLEO--CONF--96--05 (1996).

\end{thebibliography}

\end{document}